\documentclass[12pt,twoside]{article}
\usepackage{amssymb}
\usepackage{amsmath}
\usepackage{latexsym}
\usepackage{longtable}
\usepackage{epsfig}
\usepackage{graphicx,bbm,psfrag}
\graphicspath{{images/}}

\begin{document}
\vspace*{2cm}
\begin{center}
{\Large {\bf Kinetics of the Bose-Einstein Condensation\bigskip\\}} {\large{
Herbert Spohn}}\bigskip\bigskip\\Physik Department and
Zentrum Mathematik,\\ TU M\"unchen,
 D-85747 Garching, Germany\\
e-mail:~{\tt spohn@ma.tum.de}
\end{center}\bigskip\bigskip
\textbf{Abstract.} We study the bosonic Boltzmann-Nordheim kinetic
equation, which describes the kinetic regime of weakly interacting
bosons with s-wave scattering only. We consider a spatially
homogeneous fluid with an isotropic momentum distribution. The issue
of the dynamical formation of a Bose-Einstein condensate 
has been studied extensively. We supply here the
completed equations of motion for the coupled system, the energy
density distribution of the normal fluid and the density of the
condensate. With this information the post-nucleation self-similar
solution is investigated in more detail than before.
\newpage

\section{Introduction}\label{sec.1}
\setcounter{equation}{0}

The dynamics of weakly interacting quantum fluids is governed, in
approximation, by the Boltzmann-Nordheim kinetic equation. There is
no \textit{a priori} restriction on either the density or the temperature of
the quantum fluid. Since, as explained in every textbook on
Statistical Mechanics, an ideal Bose fluid undergoes a transition
from a normal fluid to one with a condensate component, one would
expect that the kinetic equation retains some information on how the
condensate is formed dynamically.

In general, the appropriate kinetic equation is fairly complex and
to analyse the details of the transition will be a difficult task.
There is however one particular case which has been investigated in
considerable depth. It is assumed that the Bose fluid is spatially
homogeneous and the momentum distribution is isotropic. The Bose
particles are massive and interact only through s-wave scattering.
The distribution function $f$, as governed by the kinetic equation,
depends then only on the energy $\varepsilon$, $\varepsilon\geq 0$,
and on time $t$, $t\geq 0$. Based on theoretical considerations and
numerical simulations the following scenario has been developed for
the kinetics of the condensation process. One starts with an initial
distribution $f(\varepsilon,0)=f(\varepsilon)$ which has the density
$\rho=\int^\infty_0 d\varepsilon \sqrt{\varepsilon}f(\varepsilon)$.
$\rho$ is assumed to be supercritical, $\rho>\rho_\mathrm{c}$. Then
the solution $f(\varepsilon,t)$ of the kinetic equation has a piece
which concentrates near $\varepsilon=0$ and  nucleates the
condensate at some finite time $t_\ast$. More precisely,
\begin{equation}\label{1.1}
\sqrt{\varepsilon}f(\varepsilon,t)=\sqrt{\varepsilon}f_\mathrm{reg}
(\varepsilon,t)+n(t)\delta(\varepsilon)
\end{equation}
with $n(t)=0$ for $t<t_\ast$ and $n(t)>0$ for $t>t_\ast$. Here
$f_\mathrm{reg}$ is a density without any delta functions. In the
limit $t\to\infty$, $n(t)$ tends to $\rho-\rho_\mathrm{c}$ while the
regular piece tends to the critical Bose-Einstein distribution
$(\mathrm{e}^{\beta \varepsilon}-1)^{-1}$ with the inverse
temperature $\beta$ determined through the initial energy
$\mathsf{e}=\int^\infty_0d\varepsilon \sqrt{\varepsilon}\varepsilon
f(\varepsilon)$. Near $t_\ast$ the solution has a self-similar
structure, which will be explained below.

In our contribution we investigate, in more detail than previous
studies, the precise mechanism of how the condensate is generated
and annihilated (on the level of the Boltzmann-Nordheim kinetic
equation). In particular we obtain additional information on the
self-similar structure of the solution for $t>t_\ast$. A numerical
check of these predictions would help to further elucidate the kinetics of
the Bose-Einstein transition.

\section{The Boltzmann-Nordheim equation for the energy distribution}\label{sec.2}
\setcounter{equation}{0}

We use throughout dimensionless variables in units which minimize
the number of prefactors. The physically correct dimensions are supplied,
\textit{e.g.},
in \cite{ST}. Instead of denoting the energy by
$\varepsilon$ it will be more convenient to use the neutral $x,y,z$.
The distribution function is denoted by $f(x)$, $f\geq 0$, with
volume element $\sqrt{x}dx$, $x\geq 0$, since for
$\varepsilon=p^2/2$ one has $d^3 p=d\Omega p^2 d|p|$ and $p^2
d|p|=\sqrt{2\varepsilon}d\varepsilon$ in three dimensions. To $f$ we
associate the density $\rho$ and energy $\mathsf{e}$ as
\begin{equation}\label{2.1}
\rho(f)=\int^\infty_0 dx \sqrt{x}f(x)\,,\quad
\textsf{e}(f)=\int^\infty_0 dx \sqrt{x} x f(x)\,.
\end{equation}
Physically $\rho$ is the mass and $\mathsf{e}$
the energy per unit volume of the spatially homogeneous Bose fluid,
but we stick to the more colloquial expressions.

$f(t)$ is governed by the kinetic equation
\begin{equation}\label{2.2}
\frac{d}{dt}f(t)=\mathcal{C}_4(f(t))\,.
\end{equation}
$\mathcal{C}_4$ is the collision operator which describes the
scattering of two incoming to two outgoing particles,
the 4-wave scattering in wave turbulence. For bosons,
and under the assumptions stated in the introduction, one obtains

\begin{eqnarray}\label{2.2a}
&&\hspace{-40pt} \mathcal{C}_4 (f)(x)= \frac{1}{\sqrt{x}}
\int_{\mathcal{D}(x)} dy dz
I(x,y,z)\big(\tilde{f}(x)\tilde{f}(w) f(y) f(z)\nonumber\\
&&\hspace{120pt}-f(x)f(w)\tilde{f}(y)\tilde{f}(z)\big)\,.
\end{eqnarray}
Here $\tilde{f}(x)=1+f(x)$, $w=y+z-x$, and the domain
$\mathcal{D}(x)=\{y,z|y\geq 0, z\geq 0, x\leq y+z\}$ for $x\geq 0$.
The integral kernel $I$ results from working out the
$\delta$-functions for energy and momentum conservation of the full
3D collision operator, where $x,w$ are the energies of the incoming
and $(y,z)$ the energies of the outgoing particles, see \cite{ST,Ca}
for details. $I$ is defined by
\begin{equation}\label{2.3}
I(x,y,z)=
 \left\{
  \begin{array}{ll}
    \sqrt{w} & \hbox{for } y\leq x \,, z \leq x\,, 0\leq w\,, \\
    \sqrt{x} & \hbox{for } x\leq y \,, x \leq z\,,\\
    \sqrt{z} & \hbox{for } x\leq y \,, z \leq x\,,\\
    \sqrt{y} & \hbox{for } y\leq x \,, x \leq z\,.
  \end{array}
\right.
\end{equation}

If $\rho(f)<\infty$, $\mathsf{e}(f)<\infty$ and if $f(t)$ is
bounded, then one can easily work out that
\begin{equation}\label{2.4}
   \frac{d}{dt}\rho(f(t))=\int^\infty_0 dx \sqrt{x}\mathcal{C}_4
(f(t))(x)=0\,,
\end{equation}
\begin{equation}\label{2.5}
   \frac{d}{dt}\textsf{e}(f(t))=\int^\infty_0 dx \sqrt{x}x \mathcal{C}_4
(f(t))(x)=0\,.
\end{equation}
Thus mass and energy are conserved. As we will see, if 
 $f$ which diverges at $x=0$, the conservation laws may break down.

On the kinetic level the entropy per unit volume is the one
of a non-interacting Bose fluid and hence given by
\begin{equation}\label{2.6}
   s(f)=\int^\infty_0dx \sqrt{x}\big(\tilde{f}(x)\log \tilde{f}(x)
-f(x)\log f(x)\big)\,.
\end{equation}
{}From (\ref{2.2}) it follows that
\begin{equation}\label{2.7}
  \frac{d}{dt}s(f(t))=\sigma(f(t))
\end{equation}
with the entropy production
\begin{equation}\label{2.8}
  \sigma(f)=\int_{0\leq x,y,z<\infty,x\leq y+z} dx dy dz
I(x,y,z)(A-B)\log (A/B)\,,
\end{equation}
\begin{equation}\label{2.8a}
 A=\tilde{f}(x)\tilde{f}(w) f(y) f(z)\,,\quad B=f(x) f(w)
\tilde{f}(y) \tilde{f}(z)\,.
\end{equation}
Clearly $\sigma(f)\geq 0$.

For a stationary solution $\sigma(f)=0$. Introducing
$\psi=\log(\tilde{f}/f)$, the condition $\sigma(f)=0$ is equivalent
to $\psi$ being a collisional invariant, to say
\begin{equation}\label{2.9}
 \psi(x)+\psi(y+z-x)=\psi(y)+\psi(z)
\end{equation}
for all $x,y,z\geq 0$ such that $x\leq y+z$. (\ref{2.9}) admits as
only solutions the affine functions $\psi(x)=c_0+c_1 x$. Together with
the condition $f\geq 0$ this yields as stationary solutions of the
Boltzmann-Nordheim equation the Bose-Einstein distributions
\begin{equation}\label{2.10}
 f_{\beta,\mu}(x)=(\mathrm{e}^{\beta(x-\mu)}-1)^{-1}\,,\quad
\beta>0\,,\mu\leq 0\,,
\end{equation}
parametrized by the inverse temperature $\beta$ and the chemical
potential $\mu$.

At given $\beta$ the maximal density is
\begin{equation}\label{2.11}
\rho_\mathrm{c}=\int^\infty_0 dx \sqrt{x}(\mathrm{e}^{\beta
x}-1)^{-1} =\beta^{-3/2}\rho_0\,,\quad \rho_0=\int ^\infty_0 dx
\sqrt{x}(\mathrm{e}^x-1)^{-1}
\end{equation}
and correspondingly the maximal energy is
\begin{equation}\label{2.11a}
\mathsf{e}_\mathrm{c}=\int^\infty_0 dx \sqrt{x}(\mathrm{e}^{\beta
x}-1)^{-1} x=\beta^{-5/2} \mathsf{e}_0\,,\quad
\mathsf{e}_0=\int^\infty_0 dx \sqrt{x}(\mathrm{e}^x-1)^{-1} x\,.
\end{equation}
Hence there is the critical line
\begin{equation}\label{2.11b}
(\mathsf{e}_\mathrm{c}/\mathsf{e}_0)=(\rho_\mathrm{c}/\rho_0)^{5/3}
\end{equation}
which divides the $(\rho,\mathsf{e})$ quadrant into the domain
$D_\mathrm{nor}$ of normal fluid and the domain $D_\mathrm{con}$ with
some fraction of condensate. If, for given initial $f$,
$(\rho(f),\mathsf{e}(f))\in D_\mathrm{con}$, then
$\rho(f)>\rho_\mathrm{c}$ and $f(t)$ has no stationary solution to
approach in the limit as $t\to \infty$. Based on equilibrium
statistical mechanics, it is thus natural to extend the class of
stationary solutions to
\begin{equation}\label{2.12}
\sqrt{x}f_{\beta,\mu}(x)+n_\mathrm{con}\delta(x)\,,\quad
n_\mathrm{con}\geq 0\,,
\end{equation}
with the condition that for $n_\mathrm{con}>0$ necessarily $\mu=0$.
The density of this distribution is denoted by
$\rho(\beta,\mu)=\rho(f_{\beta,\mu})$ for $\mu<0$ and by
$\rho(\beta,0,n_\mathrm{con})=\rho_\mathrm{c}+n_\mathrm{con}$ for
$\mu=0$. The condensate has zero energy. With this extension there
is a one-to-one relation $\Phi$ between $(\rho,\mathsf{e})$ and
$(\beta,\mu,n_\mathrm{con})$.

If in equilibrium there is a $\delta$-function at $x=0$, it seems
natural to allow for such a singular contribution also in the
dynamics. The precise formulation will be the task of the next
section. In fact, Lu \cite{Lu05} proves that, for
any $f$ with $\rho(f)< \infty$, $\mathsf{e}(f)<\infty$, in the long
time limit the solution $f(t)$ converges to (\ref{2.12}) with the
parameters $(\beta,\mu,n_\mathrm{con})$ determined through the map
$\Phi$ from the initial data. The precise statement requires some
preparations and is therefore deferred to Appendix C.

\section{Coupled equations for normal fluid and condensate}\label{sec.3}
\setcounter{equation}{0}

If there is a condensate at time $t$, the full distribution function
is
\begin{equation}\label{3.1}
   \sqrt{x}f_{\mathrm{tot}}(x,t)=\sqrt{x}f(x,t)+n(t)\delta(x)\,.
\end{equation}
We adopt the convention that $f$ always denotes a continuous
function on $\mathbb{R}_+=(0,\infty)$ such that $\rho(f)<\infty$,
$\textsf{e}(f)<\infty$ and with a possible divergence at $x=0$. We
postulate that (\ref{2.2}), (\ref{2.2a}) remain valid when $f$ is
substituted by $f_\mathrm{tot}$. Inserting $f_\mathrm{tot}$ in
(\ref{2.2a}) yields products of the form $fff$, $\delta ff$,
$\delta\delta f$, $\delta\delta\delta$, $ff$, $\delta f$, and
$\delta\delta$. Using the continuity of the integral kernel $I$, it
can be shown that the contribution of products with more than one
$\delta$ vanishes \cite{ST,Lu08}. Therefore (\ref{2.2}) turns into a
coupled system of equations for $f$ and $n$, apparently first
derived in \cite{ST1}, which reads
\begin{equation}\label{3.2}
 \frac{d}{dt}f=\mathcal{C}_4(f)+n\mathcal{C}_3(f)\,,
\end{equation}
\begin{equation}\label{3.3}
 \frac{d}{dt}n=-n\int^\infty_0 dx
\sqrt{x}\mathcal{C}_3(f)(x)\,.
\end{equation}
The collision operator $\mathcal{C}_3$ is familiar from 3-wave
interactions. Here it arises since one of the collision
partners is at rest. Using isotropy and integrating over the
momentum and energy delta functions, one obtains,
\begin{eqnarray}\label{3.4}
&&\hspace{-20pt} \mathcal{C}_3 (f)(x)= \frac{2}{\sqrt{x}} \int^x_0
dy \big(\tilde{f}(x)f(x-y)f(y)-f(x)\tilde{f}(x-y) \tilde{f}(y)\big)\nonumber\\
&&\hspace{40pt}+\frac{4}{\sqrt{x}} \int^\infty_x dy
\big(\tilde{f}(x)\tilde{f}(y-x)f(y)-f(x)f(y-x) \tilde{f}(y)\big)\,,
\end{eqnarray}
see \cite{ST} for details.

Of course, to properly justify (\ref{3.2}), (\ref{3.3}), one has to
go back to the microscopic model of a fluid of weakly interacting
bosons. One imposes an initial quasifree state,
$\langle\cdot\rangle$, such that
\begin{eqnarray}\label{3.4a}
&&\hspace{15pt} \langle a(k)\rangle= \alpha\delta(k)\,, \quad
\langle a(k)a(k')\rangle= \delta(k)\delta(k')\alpha^2\,,
\nonumber\\
&&\hspace{15pt}\langle a(k)^\ast a(k')\rangle=
\delta(k-k')\big(|\alpha|^2\delta(k)+f(k^2)\big)\,,
\end{eqnarray}
where $a(k)$, $a(k)^\ast$ are the bosonic annihilation and creation
operators labeled by momentum $k\in\mathbb{R}^3$.
$\alpha\in\mathbb{C}$ and $|\alpha|^2$ is the condensate density.
The state $\langle\cdot\rangle$ is invariant under spatial
translations. If the BBGKY hierarchy is quasifreely truncated at the
sixth order, then at least formally one arrives at the coupled
system (\ref{3.2}), (\ref{3.3}).

$\mathcal{C}_3$ conserves energy, but not mass. By
considering the H-Theorem as in (\ref{2.7}), one concludes that
$\sigma(f)=0$ is equivalent to $\psi=\log (\widetilde{f}/f)$
being a collisional invariant, in the sense that
\begin{equation}\label{3.4b}
\psi(x)+\psi(y)=\psi(x+y)
\end{equation}
for all $x,y\geq 0$. Clearly, $\psi$ must be linear. Hence the
stationary solutions for $\mathcal{C}_3$ are given by $f_{\beta,0}$
with $\beta>0$, where $\beta$ is determined by the initial energy
$\mathsf{e}(f)$.

The coupled system (\ref{3.2}), (\ref{3.3}) has an obvious defect.
If $n(0)=0$, then $n(t)=0$ for all $t$, no condensate is nucleated.
A second defect lies in the definite sign of the term on the right
hand side of (\ref{3.3}). If $f$ is bounded, then $\int^\infty_0
dx\sqrt{x}\mathcal{C}_3(f)(x)\geq 0$, as will be shown below. Hence,
if present initially, the condensate density is monotone decreasing.
The derivation of (\ref{3.2}), (\ref{3.3}) provides no immediate
indication of how the condensate is generated. Only, since physically the total
mass is conserved, the mass lost by the normal component has
to be identified with the mass of the condensate.

In view of (\ref{2.4}) valid for bounded $f$'s a natural guess is
that the loss of mass for (\ref{3.2}) is linked to the divergence of
$f$ at $x=0$. The relevant properties are stated in two
propositions, which will be
proved in Appendix A.\medskip\\
\textbf{Proposition 1.} \textit{Let $f$ be continuous on $\mathbb{R}_+$, $\rho(f)<
\infty$, $\mathsf{e}(f)< \infty$, and let
\begin{equation}\label{3.5}
 \lim_{x\to 0} x^{7/6} f(x)=b,\quad b \geq 0\,.
\end{equation}
Then
\begin{equation}\label{3.6}
 \lim_{\delta\to 0}\int^\infty_\delta dx \sqrt{x}\mathcal{C}_4
(f)(x)=-\Gamma_4 b^3\,.
\end{equation}
Here $\Gamma_4\cong 3.05$ and the defining integral is provided in
(\ref{A.11})}.
\medskip\\
\textbf{Proposition 2.} \textit{Let $f$ be continuous on $\mathbb{R}_+$, $\rho(f)<
\infty$, $\mathsf{e}(f)<\infty$, and let
\begin{equation}\label{3.7}
    \lim_{x\to 0} x f(x)=a, \quad a \geq 0\,.
\end{equation}
Then
\begin{equation}\label{3.8}
 \lim_{\delta\to 0}\int^\infty_\delta dx \sqrt{x}\mathcal{C}_3
(f)(x)=-\Gamma_3 a^2+ 2\int^\infty_0 dx x f(x)\,.
\end{equation}
Here $\Gamma_3 =- 2\int^1_0 dx x^{-1}\log
(1-x)=\pi^2/3$.}\medskip\\
\textit{Remark:} In
agreement with the fact that the condensate has zero energy, under
the condition (\ref{3.5}) energy is still conserved in the sense
that
\begin{equation}\label{3.9}
 \lim_{\delta\to 0}\int^\infty_\delta dx \sqrt{x} x
\mathcal{C}_j(f)(x)=0\quad \textrm{for } j=3,4\,.
\end{equation}
{\hspace*{\fill}$\diamond$}\\
 In (\ref{3.8}) there is a loss of mass as $\Gamma_3 a^2$ and a
 gain given by the integral expression. We interpret this as the
 corresponding loss/gain of the condensate. Therefore (\ref{3.2}) and
(\ref{3.3}) are improved to
\begin{equation}\label{3.10}
 \frac{d}{dt}f(t)=\mathcal{C}_4(f(t))+n(t)\mathcal{C}_3(f(t))\,,
\end{equation}
\begin{equation}\label{3.11}
 \frac{d}{dt}n(t)= n(t)\big(\Gamma_3 a(t)^2
-2\int^\infty_0 dx x f(x,t)\big)\,.
\end{equation}
Note that for $f_{\beta,0}$ one has
\begin{equation}\label{3.12}
a=\beta^{-1}\,,\quad 2 \int^\infty_0 dx x
f_{\beta,0}(x)=\beta^{-2}\Gamma_3\,.
\end{equation}
Thus the Bose-Einstein distributions of (\ref{2.12}) are indeed
stationary solutions for (\ref{3.10}), (\ref{3.11}).

In (\ref{3.11}) we did not include the term $\Gamma_4 b(t)^3$, since
$b>0$ implies $a=\infty$. Mass is transferred between normal fluid and condensate
only through $n
\mathcal{C}_3$. Thus it seems that the coupled system
(\ref{3.10}), (\ref{3.11}) still
has the defect that, in case $n(t)=0$, no condensate can be nucleated. We will see
in the next section, how this objection is met.

Tentatively we assume that a solution $f(t),n(t)$ to
(\ref{3.10}),
(\ref{3.11}) has the properties:\smallskip\\
(i) $f(t)\geq 0$, $n(t)\geq 0$ are continuous in $t$,
$\rho(f(t))<\infty$, $\textsf{e}(f(t))<\infty$.\\
(ii) $f(t):\mathbb{R}_+\to \overline{\mathbb{R}_+}$ is continuous
and
$\lim_{x\to 0} x f(x,t)=a(t)$ exists with  $a(t)<\infty$.

\section{Scaling theory close to nucleation of condensate}\label{sec.4}
\setcounter{equation}{0}

We note that if $n(t_0)>0$ for some $t_0$, then $n(t)>0$ for all
$t>t_0$. Thus, if $n(0)=0$, there must be a first time, $t_\ast$,
such that $n(t)=0$ for $t<t_\ast$ and $n(t)>0$ for $t>t_\ast$. Of
course, depending on the initial $f$, one could have
$t_\ast=\infty$. In this section we plan to study the solution close
$t_\ast$, the time of nucleation of the condensate. We mostly follow
\cite{JPR}, but the post-nucleation seems to be novel.\medskip\\
\textit{(i) $n(0)=0$, $t_\ast<\infty$, pre-nucleation of the
condensate.}

For the scaling theory the quadratic terms of $\mathcal{C}_4$ can be
neglected. Physically this corresponds to the semiclassical
approximation. Since $n(t)=0$ for $t<t_\ast$, we have to study
\begin{equation}\label{4.1}
\frac{d}{dt}f=\mathcal{C}_{4,\mathrm{sc}}(f)
\end{equation}
with the semiclassical collision operator
\begin{eqnarray}\label{4.2}
&&\hspace{-25pt} \mathcal{C}_{4,\mathrm{sc}} (f)(x)=
\frac{1}{\sqrt{x}} \int_{\mathcal{D}(x)}
dy dz I(x,y,z)\big(f(x)f(y)f(z)+f(w)f(y)f(z)\nonumber\\
&&\hspace{95pt}-f(x)f(w)f(y)-f(x)f(w)f(z)\big)\,.
\end{eqnarray}

Close to $x=0$ the scaling ansatz is
\begin{equation}\label{4.3}
f(x,t_\ast-t)=\frac{1}{\tau(t)^\nu}\phi_-\big(\frac{x}{\tau(t)}\big)\,,\quad
0<t\ll 1\,.
\end{equation}
Inserting in (\ref{4.1}), (\ref{4.2}) yields
\begin{equation}\label{4.4}
\tau(t)^{-\nu-1}\frac{d}{dt}\tau(t)\big(\nu\phi_-(x)+x\phi_-'(x)\big)
=\tau^{-3\nu+2}\mathcal{C}_{4,\mathrm{sc}}(\phi_-)(x)\,.
\end{equation}
We fix the time scale by setting $\tau^{2\nu-3}\dot{\tau}=1$. Then,
for $\nu>1$,
\begin{equation}\label{4.5}
\tau(t)=\big(2(\nu-1)t\big)^{1/2(\nu-1)}\,.
\end{equation}
The scaling function $\phi_-$ becomes then the solution of the nonlinear
eigenvalue problem
\begin{equation}\label{4.6}
\nu\phi_-(x)+x\frac{d}{dx}\phi_-(x)=\mathcal{C}_{4,\mathrm{sc}}(\phi_-)(x)\,.
\end{equation}

For $\tau\ll x\ll 1$ the solution in (\ref{4.3}) is assumed to be
frozen at a definite power law. This yields the right boundary condition
\begin{equation}\label{4.7}
\phi_-(x)\cong x^{-\nu}\quad \textrm{for } x\to\infty\,.
\end{equation}
For small $x$ the, conventional, left boundary condition is
\begin{equation}\label{4.8}
\phi_-(x)\to \phi_0\quad \textrm{for } x\to 0\,.
\end{equation}
In the limit $x\to 0$, (\ref{4.6}) then yields, noting that the
terms in $\mathcal{C}_{4,\mathrm{sc}}(\phi_+)$ proportional
to $\phi_0$ cancel each other,
\begin{equation}\label{4.9}
\nu \phi_0=\int^\infty_0 dy \int^\infty_0 dz
\phi_-(y)\phi_-(z)\phi_-(y+z)\,,
\end{equation}
which is consistent in the sense that both sides have the same sign.

$\nu$ and $\phi_0$ are free parameters. To have condensation at all,
by Proposition 1 necessarily
\begin{equation}\label{4.9a}
\nu\geq 7/6\,.
\end{equation}
On the other hand to have a finite mass in the interval $0\leq x\leq
\tau(t)$ yields the upper bound
\begin{equation}\label{4.9b}
\nu< 3/2\,.
\end{equation}
Because of slow decay, the scaling function $\phi_-$ has infinite mass and
energy. It would be nice, if (\ref{4.6}) would determine a unique
value of $\nu$. But very little is known in this direction.

Numerically one finds that
\begin{equation}\label{4.10}
\nu=1.234
\end{equation}
by fitting the power law of $f(x,t_\ast-t)$ for $x>\tau$, see \cite{ST1,LLPR}.
One also finds
$f(0,t_\ast - t)\cong  \tau(t)^{-\nu}\phi_0$ consistent with the right boundary
condition (\ref{4.8}). We refer to \cite{JPR} for additional
information on the eigenvalue
equation (\ref{4.6}).

At the onset of nucleation one has $f(x,t_\ast)= x^{-\nu}$, which is outside
the space of solutions. As we will argue next, for any $t > t_\ast$
the solution falls back into its proper space.\medskip\\
\textit{(ii) $n(0)=0$, $t_\ast<\infty$, post-nucleation of the
condensate.}

As for $t<t_\ast$ we can work with the semiclassical approximation,
which amounts to
\begin{equation}\label{4.11}
\frac{d}{dt}f=\mathcal{C}_{4,\mathrm{sc}}(f)+
n\mathcal{C}_{3,\mathrm{sc}}(f)\,,
\end{equation}
\begin{equation}\label{4.12}
\frac{d}{dt}n= n \Gamma_3 a^2 \,.
\end{equation}
The 3-wave semiclassical collision operator reads
\begin{eqnarray}\label{4.13}
&&\hspace{-25pt} \mathcal{C}_{3,\mathrm{sc}} (f)(x)=
\frac{2}{\sqrt{x}} \int^x_0 dy \{f(x-y)f(y)-f(x)f(x-y)-f(x)f(y)\}\nonumber\\
&&\hspace{43pt}+\frac{4}{\sqrt{x}} \int^\infty_x dy
\{f(x)f(y)+f(y)f(y-x)-f(x)f(y-x)\}\,.
\end{eqnarray}
We impose the scaling ansatz for $x$ close to 0 as
\begin{equation}\label{4.14}
f(x,t_\ast+t)=\frac{1}{\tau(t)^{\widetilde{\nu}}}\phi_+\big(\frac{x}{\tau(t)}\big)\,,\quad
0<t\ll 1\,.
\end{equation}
The boundary condition at infinity is
\begin{equation}\label{4.15}
\phi_+(x)=x^{-\widetilde{\nu}}\quad \textrm{for } x\to\infty\,.
\end{equation}

If $f$ denotes the continuation of the scaling solution from \textit{(i)},
then one should set 
\begin{equation}\label{4.14a}
\widetilde{\nu}=\nu\,,
\end{equation}
since for a very short
time span and away from $x=0$ the distribution function is frozen. On the
other hand a general $\widetilde{\nu}$ is of interest, since one can
set $t_\ast =0$ and impose by hand some power law
for the intial distribution, i.e. one sets
\begin{equation}\label{4.15a}
f(x)=x^{-\widetilde{\nu}} h(x)\quad \textrm{for } t=0
\end{equation}
with a cutoff function $h$ so to have bounded $\rho(f)$,
$\mathsf{e}(f)$. Here necessarily
\begin{equation}\label{4.15b}
1\leq\widetilde{\nu}< 3/2\,,
\end{equation}
the lower bound being required for nucleation, the upper bound for
$\rho(f)<\infty$.

The condensate density is assumed to scale as
\begin{equation}\label{4.16}
n(t)=(t/t_0)^\gamma\,.
\end{equation}
Inserting in (\ref{4.12}) yields
\begin{equation}\label{4.16b}
\gamma t^{-1}=\Gamma_3 a(t)^2\,.
\end{equation}
To ensure $a > 0$, we have to impose the left boundary condition
\begin{equation}\label{4.17}
\phi_+(x)=\frac{a_0}{x}\quad
\textrm{for } x\to 0\,,\; a_0>0\,.
\end{equation}
Then
\begin{equation}\label{4.18a}
a(t)=\lim_{x\to 0} x
\tau^{-\widetilde{\nu}}\phi_+(\frac{x}{\tau})=\tau^{-\widetilde{\nu}+1}
a_0\,,
\end{equation}
which implies
\begin{equation}\label{4.24}
\tau(t)^{-\widetilde{\nu}+1}=\frac{1}{a_0}(\frac{\gamma}{\Gamma_3})^{1/2}
t^{-1/2}\,.
\end{equation}

Next we insert the scaling ansatz (\ref{4.14}), (\ref{4.16}) in
(\ref{4.11}). Multiplying both sides by $\tau^{3\nu-2}$ one arrives
at
\begin{equation}\label{4.25}
-\tau^{2\nu-3}\dot{\tau}(\widetilde{\nu}\phi_+ +
x\phi'_+)=\mathcal{C}_{4,\mathrm{sc}}(\phi_+) +(t/t_0)^\gamma
\tau^{\widetilde{\nu}-3/2}\mathcal{C}_{3,\mathrm{sc}}(\phi_+)\,.
\end{equation}
The prefactor on the left turns out to be independent of $t$.
Balancing the prefactor of $\mathcal{C}_{3,\mathrm{sc}}$ yields
\begin{equation}\label{4.26}
\gamma=\frac{3-2\widetilde{\nu}}{4(\widetilde{\nu}-1)}\,.
\end{equation}
$\gamma>0$ provided $1<\widetilde{\nu}< 3/2$ consistent with
(\ref{4.15b}). We conclude that the scaling function $\phi_+$ satisfies
\begin{equation}\label{4.27}
-K_1 (\widetilde{\nu}\phi_+ +
x\phi'_+)=\mathcal{C}_{4,\mathrm{sc}}(\phi_+)+K_2
\mathcal{C}_{3,\mathrm{sc}}(\phi_+)
\end{equation}
with the boundary conditions (\ref{4.15}) and (\ref{4.17}) and the two
positive constants
\begin{equation}\label{4.28}
K_1=a^2_0 T_3(\tfrac{3}{2}-\widetilde{\nu})^{-1}\,,\quad
K_2=\big(\frac{1}{a^2_0 t_0}\frac{\gamma}{\Gamma_3}\big)^\gamma\,.
\end{equation}
$a_0$ and $t_0$ are free parameters.

There is one important consistency check. In (\ref{4.27}) we take
the limit $x\to 0$. The left hand side behaves as $-K_1
a_0(\widetilde{\nu}-1)/x$. For $\mathcal{C}_{4,\mathrm{sc}}$ we use,
see Appendix A,
\begin{equation}\label{4.29}
\lim_{x\to 0}x\mathcal{C}_{4,\mathrm{sc}}(\phi_+)(x)= (a_0)^3
\Gamma_{4,1}\,,\quad \Gamma_{4,1}= \int^\infty_1 dy \int^\infty_1 dz
\big((y+z-1)yz\big)^{-1}\,.
\end{equation}
Numerically $ \Gamma_{4,1} = 1.645$. Clearly, one  cannot balance the negative
$-K_1 a_0(\widetilde{\nu}-1)$. Therefore,
$\mathcal{C}_{3,\mathrm{sc}}$ must come into play. For small
arguments the leading terms in the integrand of
$\mathcal{C}_{3,\mathrm{sc}}$ cancel and, to achieve the $1/x$
singularity, we have to assume the subleading behavior
\begin{equation}\label{4.30}
\phi_+ (x)^{-1} = a_0 x +a_1 x^{3/2}\quad \textrm{for } x\to 0\,.
\end{equation}
Then, see Appendix A,
\begin{eqnarray}\label{4.31}
&&\hspace{-20pt}\lim_{x\to
0}x\mathcal{C}_{3,\mathrm{sc}}(\phi_+)(x)= - (a_0)^3 a_1
\Gamma_{3,1}\,,\nonumber\\
&&\hspace{-20pt}\Gamma_{3,1}= - \int^1_0 du \frac{1}{u(1-u)}
(2-4u^{-1/2})(1-u^{3/2}-(1-u)^{3/2})\,.
\end{eqnarray}
Numerically $\Gamma_{3,1}= 5.56$. We arrive at
\begin{equation}\label{4.32}
-K_1 a_0(\widetilde{\nu}-1)=a^3_0 \Gamma_{4,1} - K_2 a^3_0 a_1
\Gamma_{3,1}\,,
\end{equation}
which determines $a_1$. Thus $\phi_+$ has a subleading behavior as
$-a_1 x^{-1/2}$ for $x\to 0$. By the same procedure one could, in
principle, compute the next order corrections to $\phi_+$.

At first sight the nonlinear eigenvalue problems (\ref{4.6}) and
(\ref{4.27}) look rather similar, except for a further collision
operator and the change of sign on the left. On physical grounds we
expect however a very different solution behavior. (\ref{4.6})
should have a single eigenvalue, or possibly a discrete set of 
eigenvalues \cite{JPR}, while (\ref{4.27}) should have for
any $\widetilde{\nu}$ with $1<\widetilde{\nu}<3/2$ a solution
satisfying the boundary conditions (\ref{4.15}) and (\ref{4.17}).

Physically the most interesting, and accessible, prediction is the
exponent $\gamma$ of (\ref{4.26}) which governs the initial increase
in the condensate density. If in (\ref{4.15a}). one sets
$7/6\leq\widetilde{\nu}< 3/2$, then, according to Proposition 1,
$\mathcal{C}_4$ provides a mechanism for nucleation. Immediately,
i.e. for any $t>0$, the divergence at $x=0$ drops to $a(t)/x$,
$a(t)$ from (\ref{4.16b}). For $\widetilde{\nu}=\nu\cong 1.234$
numerical simulations are available \cite{LLPR} and yield
$\gamma=0.571$ in good agreement with the prediction (\ref{4.26}).
At $\widetilde{\nu}=7/6$, (\ref{4.26}) results in $\gamma=1$.
Aspects of this case have been established mathematically by
Escobedo \textit{et al.} and we state their results in Appendix B.
Also the numerical solutions in \cite{ST,ST1} clearly show a linear
initial increase in the condensate density and the rapid switch from
the initial divergence $x^{-7/6}$ to the slower $x^{-1}$. For
$1<\widetilde{\nu}< 7/6$ and $n(0)=0$, no mechanism for nucleation
at time $t=0$ is available. (Of course, at some later time the
scenario as described under \textit{(i)} may set in). Thus to verify the
prediction (\ref{4.26}) one would have to start with some small
$n(0)$, so to set $\mathcal{C}_3$ in action.

\section{Conclusions}\label{sec.5}
\setcounter{equation}{0}

In the ``naive'' picture of the condensation process the
distribution function $f$ develops a $\delta$-like concentration
close to the origin, either at some finite time or in the long time
limit. The Boltzmann-Nordheim kinetic equation tells a different
story, however: If there is sufficient mass accumulated near the
origin, the solution develops explosively a $1/x$ singularity at 0
(which is integrable for $\sqrt{x}dx$). Once the singularity is
formed, mass of the normal fluid can be channeled into the condensate.
On the other hand the condensate is annihilated at a rate
$2\int^\infty_0 dx x f(x,t)$. For long times both processes balance
so to approach the equilibrium condensate density.

\begin{appendix}
\section{Appendix: Proofs of Propositions 1, 2 and Eqs. (\ref{4.29}), (\ref{4.31}).}\label{sec.A}
 \setcounter{equation}{0}

We start with the slightly easier proof of Proposition 2.\\
\textit{Proof of Proposition 2:}
We have to study the limit $\delta \to 0$ of
\begin{eqnarray}\label{A.1}
&&\hspace{-8pt} 2\int^\infty_\delta dx \int^x_0 dy
\big(\{ f(x-y) f(y)-f(x) f(x-y)- f(x) f(y)\}-f(x)\big)\nonumber\\
 &&\hspace{30pt}+4 \int^\infty_\delta dx \int^\infty_x dy
\big(\{ f(x) f(y)+f(y-x) f(y)- f(x) f(y-x)\}+f(y)\big)\nonumber\\
 &&\hspace{30pt}=A_1(\delta)+
A_2(\delta)+A_3(\delta)\,,
\end{eqnarray}
where $A_1(\delta)$ is the integral with the first curly bracket,
$A_2(\delta)$ the one with the second curly bracket, and
$A_3(\delta)$ is the sum of the two integrals linear in $f$.

We consider $A_1(\delta)$. Each summand is linearly transformed such
that the integrand is of product form, $f(x)f(y)$. For $f(x-y)f(y)$
the new domain of integration is $\{x,y\geq 0|\delta\leq x+y\}$, for
$f(x)f(x-y)$ the new domain of integration is $\{x,y\geq 0|x\leq
y\,,\;\delta\leq y\}$, and for $f(x)f(y)$ the domain of integration
is $\{x,y\geq 0|\delta\leq x\,,\;y\leq x\}$. Altogether
\begin{equation}\label{A.2}
A_1(\delta)=2\int^\delta_0 dx \int^\delta_{\delta-x} dy f(x) f(y)\,.
\end{equation}
By continuity one has the bounds $a_-(\delta)x^{-1}\leq f(x)\leq
a_+(\delta)x{-1}$ valid for $0\leq x\leq\delta$ with $\lim_{\delta\to
0}a_\pm(\delta)=a$. Inserting in (\ref{A.2}) yields
\begin{equation}\label{A.3}
\Gamma_3 a_-(\delta)^2\leq A_1(\delta)\leq \Gamma_3 a_+(\delta)^2\,,
\end{equation}
hence
\begin{equation}\label{A.3a}
\lim_{\delta\to 0} A_1(\delta)=\Gamma_3 a^2\,.
\end{equation}

Next we study $A_2(\delta)$ by the same technique. For $f(x)f(y)$
the domain of integration is $\{x,y\geq 0|\delta\leq x<\infty,x\leq
y\}$, for $f(y-x)f(y)$ the new domain of integration is $\{x,y\geq
0| 0\leq x< \infty, x+\delta\leq y\}$, and for $f(x)f(y-x)$ the new
domain of integration is $\{x,y\geq 0|\delta\leq x<\infty\}$.
Altogether, using the symmetry of the integrand, one obtains
\begin{equation}\label{A.4}
A_2(\delta)=-4\int^\infty_\delta dx \int^x_{x-\delta} dy f(x)
f(y)\,.
\end{equation}
We split the $x$ integration into $\delta\leq x <\delta_1$ and
$\delta_1\leq x<\infty$, with $\delta<\delta_1$. For the second
interval
\begin{equation}\label{A.5}
\lim_{\delta\to 0}\int^\infty_{\delta_1} dx \int^x_{x-\delta} dy
f(x) f(y)=0
\end{equation}
by dominated convergence, since $\int^\infty_{\delta_1} dx
f(x)<\infty$ and $\int^x_{x-\delta} dy f(y)\to 0$ for $\delta\to 0$
with $\delta_1\leq x$. For the interval $\delta\leq x\leq \delta_1$
we use the bounds $a_-(\delta_1)x^{-1}\leq f(x)\leq
a_+(\delta_1)x^{-1}$ valid for $0\leq x\leq\delta_1$ with
$\lim_{\delta_1\to 0}a_\pm(\delta_1)=a$. Letting $\delta\to 0$ and
subsequently $\delta_1\to 0$, we conclude
\begin{equation}\label{A.5a}
\lim_{\delta\to 0} A_2(\delta)=-2\Gamma_3 a^2\,.
\end{equation}

Finally for the linear terms we use that
\begin{equation}\label{A.6}
\int^\infty_0 dx f(x) x<\infty\,.
\end{equation}
Therefore
\begin{equation}\label{A.7}
\lim_{\delta\to 0}\big(A_1(\delta)+A_2(\delta)+A_3(\delta)\big)=
-\Gamma_3 a^2 +2 \int^\infty_0 dx x f(x)\,.
\end{equation}
{\hspace*{\fill}$\Box$}\medskip\\
\textit{Proof of Proposition 1:} We have to study the limit
$\delta\to 0$ of
\begin{eqnarray}\label{A.8}
&&\hspace{-8pt} \int^\infty_\delta dx \int_{\mathcal{D}(x)} dy dz
I(x,y,z)\big(\{ f(x) f(y) f(z) + f(w) f(y) f(z) - f(x) f(w) f(y)\nonumber\\
 &&\hspace{30pt}- f(x) f(w) f(z) \}+f(y) f(z)- f(x) f(w)\big)\nonumber\\
 &&\hspace{30pt}=B_1(\delta) + B_2(\delta)\,,
\end{eqnarray}
where $B_1(\delta)$ is integral with the curly brackets and
$B_2(\delta)$ is the one quadratic in $f$.

We consider first $B_1(\delta)$. For each summand we make a linear
change of variables to the product form $f(x) f(y) f(z)$. For $f(x)
f(y) f(z)$ the domain integration is $\{x,y,z\geq 0|\delta\leq
x,x\leq y+z\}$. For $f(w)f(y)f(z)$ the new domain of integration is
$\{x,y,z\geq 0|x+\delta\leq y+z\}$. Using the symmetry of the
integrand, for  $f(x)f(w)f(y)$ the new domain of integration is
$\{x,y,z\geq 0|x<y+z, z>\delta\}$ and for $f(x)f(w)f(z)$ the new one
is $\{x,y,z\geq 0|x<y+z, y>\delta\}$. Under these transformations
the integral kernel $I$ is not altered. Adding the four terms yields
\begin{eqnarray}\label{A.9}
&&\hspace{-16pt} B_1(\delta)= -2  \int^\delta_0 dx
\int^{x+\delta}_\delta dy \int^{x-y+\delta}_0 dz \sqrt{z} f(x) f(y) f(z) \nonumber\\
 &&\hspace{30pt}+ \int^\delta_0 dx
\int^\delta_x dy \int^\delta_{x-y+\delta} dz \sqrt{x} f(x) f(y) f(z)\nonumber\\
 &&\hspace{30pt} - \int^\delta_0 dx
\int^\infty_\delta dy \int^\infty_\delta dz \sqrt{x} f(x) f(y) f(z) \nonumber\\
 &&\hspace{30pt}- \int^\infty_\delta dx
\int^x_\delta dy \int^{x-y+\delta}_{\max(x-y,\delta)} dz \sqrt{y+z-x} f(x) f(y) f(z) \nonumber\\
 &&\hspace{30pt} + 2\int^\infty_\delta dx
\int^\delta_0 dy \int^\infty_{x-y+\delta} dz \sqrt{y} f(x) f(y)
f(z)\,.
\end{eqnarray}
As in the proof of Proposition 2, one splits the domain of the $x$
integration into $0\leq x \leq\delta_1$, $\delta_1\leq x<\infty$
with $\delta<\delta_1$. By continuity one has the bounds
$b_-(\delta_1)x^{-7/6}\leq f(x)\leq b_+(\delta_1)x^{-7/6}$ valid for
$0\leq x\leq \delta_1$ with $\lim_{\delta_1\to 0}
b_\pm(\delta_1)=b$. Since $\rho(f)+\mathsf{e}(f)<\infty$, we
conclude that in the limit $\delta\to 0$ the integral over the
interval $\delta_1\leq x<\infty$ vanishes. Inserting the bounds, the
integral over $0\leq x<\delta_1$ becomes
\begin{equation}\label{A. 10}
\Gamma_4 b_-(\delta_1)^3\leq\lim_{\delta\to 0} B_1(\delta) \leq
\Gamma_4 b_+(\delta_1)^3\,,
\end{equation}
where $\Gamma_4$ is independent of $\delta_1$ and given by
\begin{eqnarray}\label{A.11}
&&\hspace{2pt} -\Gamma_4= -2  \int^1_0 dx
\int^{1+x}_1 dy \int^{x-y+1}_0 dz  \sqrt{z}(xyz)^{-7/6} \nonumber\\
 &&\hspace{30pt}+ \int^1_0 dx
\int^1_x dy \int^1_{x-y+1} dz \sqrt{x} (xyz)^{-7/6}\nonumber\\
 &&\hspace{30pt} - \int^1_0 dx
\int^\infty_1 dy \int^\infty_1 dz \sqrt{x} (xyz)^{-7/6} \nonumber\\
 &&\hspace{30pt}- \int^\infty_1 dx
\int^x_1 dy \int^{x-y+1}_{\max(x-y,1)} dz \sqrt{y+z-x} (xyz)^{-7/6} \nonumber\\
 &&\hspace{30pt} + 2\int^\infty_1 dx
\int^1_0 dy \int^\infty_{x-y+1} dz \sqrt{y} (xyz)^{-7/6}\,.
\end{eqnarray}
The result follows by taking $\delta_1\to 0$.

Finally we discuss $B_2(\delta)$. Under our assumptions each one of
the integrals is finite separately. Therefore
\begin{equation}\label{A.12}
\lim_{\delta\to 0} B_2(\delta)= \int^\infty_0 dx
\int_{\mathcal{D}_x} dy dz I(x,y,z) \big(f(y) f(z)- f(x)
f(y+z-x)\big)=0\,,
\end{equation}
by a linear change of coordinates in the second summand.
{\hspace*{\fill}$\Box$}\medskip\\
$\Gamma_4$ cannot be computed explicitly. Numerically one finds
$\Gamma_4=3.05$. 

We supply the proofs of (\ref{4.29}) and
(\ref{4.31}). \medskip\\
\textbf{Lemma 3.} \textit{Let $f$ be continuously differentiable on
$\mathbb{R}_+$, $f \geq 0$, $\int^\infty_1 dx f(x)<\infty$, and
\begin{equation}\label{A.13}
\lim_{x\to 0} x f(x)= a_0\geq 0\,.
\end{equation}
Then
\begin{equation}\label{A.14}
\lim_{x\to 0} x \mathcal{C}_{4,\mathrm{sc}}(f)(x)= (a_0)^3
\int^\infty_1 dy \int^\infty_1 dz ((y+z-1)yz)^{-1}\,.
\end{equation}}\\
\textit{Proof:} The domains in the definition of $\mathcal{D}(x)$
are labelled by I, II, III, IV in the order of (\ref{2.3}). Each
domain will be discussed separately. \\
\textit{ad I}: All arguments are small and one can use that $f(x)=a_0/x$ for
small $x$. Then one has
\begin{eqnarray}\label{A.15}
&&\hspace{-20pt} x f(x) \frac{1}{\sqrt{x}}\int^x_0 dy \int^x_{x-y}
dz
\sqrt{w}\big(f(y) f(z) - f(w) f(y) - f(w) f(z)\big)\nonumber\\
 &&\hspace{10pt}+\frac{1}{\sqrt{x}} \int^x_0 dy \int^x_{x-y} dz
 \sqrt{w} f(w)f(y) f(z)\nonumber\\
 &&\hspace{10pt}\cong (a_0)^3\Big( -\int^1_0 dy \int^1_{1-y} dz
 \big(yz\sqrt{y+z-1}\big)^{-1}\nonumber\\
  &&\hspace{10pt}+ \int^1_0 dy \int^1_{1-y} dz (yz\sqrt{y+z-1})^{-1}\Big)=0 
\end{eqnarray}
valid for small $x$,
where we use that in the last line the integral is well defined.\\
\textit{ad II}: We first note that
\begin{equation}\label{A.17}
x f(x)\int^\infty_x dy \int^\infty_x dz \big( f(y)f(z)-f(y+z-x)
f(y)- f(y+z-x) f(z)\big)=0\,,
\end{equation}
since each term of the integrand is integrable. For the remaining
integral we choose a fixed $\delta>0$ and split the $y,z$
integration into $[x,\delta]$ and $[\delta,\infty)$. The integral
over $[\delta,\infty)\times[\delta,\infty)$ is proportional to $x$ and
the off-diagonal part is proportional to $x\log x$. Thus it remains
to study the domain $[x,\delta]$ and $[x,\delta]$. Choosing $\delta$
small and approximating $f(x)$ by $a_0/x$, one obtains
\begin{equation}\label{A.18}
(a_0)^3 x \int^\delta_x dy \int^\delta_x dz
\big((y+z-x)yz\big)^{-1}= (a_0)^3 x \int^{\delta/x}_1 dy
\int^{\delta/x}_1 dz \big((y+z-1)yz\big)^{-1}
\end{equation}
which tends to the expression in (\ref{A.14}) as $x\to 0$.\\
\textit{ad III} (and \textit{ad IV} by symmetry): We fix $\delta>0$ such that $(x
f(x)-a_0|\leq c(\delta)$ with $c(\delta)\to 0$ as $\delta\to 0$. We
split the integral over $y$ in $[0,\delta]$ and $[\delta,\infty)$.
For the first domain the argument in \textit{ad I} shows that the limit $x\to
0$ vanishes. The remaining integral is
\begin{eqnarray}\label{A.16}
&&\hspace{-20pt} \sqrt{x} \int^\infty_\delta dy \int^x_0 dz
\sqrt{z}\big(f(x) f(z) (f(y)-f(y+z-x))\nonumber\\
&&\hspace{70pt}+ f(w) f(y) f(z)- f(x) f(w) f(y)\big)\\
  &&\hspace{10pt}\cong \sqrt{x}\int^\infty_\delta dy \int^x_0 dz
 \sqrt{z}\big(a_0 x^{-1} f(z)f'(y)(x-z)+f(y)^2 a_0 z^{-1}-
 a_0x^{-1} f(y)^2\big)\,,\nonumber
\end{eqnarray}
which is of order $x$ and vanishes as $x\to 0$.
\medskip\\
\textbf{Lemma 4.} \textit{Let $f$ be once continuously differentiable on
$\mathbb{R}_+$, $f\geq 0$, $\int^\infty_1 dx f(x)<\infty$ and there
exists constants $\varepsilon,\delta$, $\varepsilon>0$, $\delta>0$,
$a_0,b_0,c_0$, such that
\begin{equation}\label{A.19}
|f(x)^{-1} -a_0 x-b_0 x^{3/2}|\leq c_0 x^{3/2+\varepsilon}
\end{equation}
for $x\in[0,\delta]$. Then
\begin{equation}\label{A.20}
\lim_{x\to 0} x \mathcal{C}_{3,\mathrm{sc}}(f)(x)= (a_0)^3 b_0
\int^1_0 du \frac{1}{u(1-u)}(2-4u^{-1/2}) (1-u^{3/2}-(1-u)^{3/2})\,.
\end{equation}}\\
\textit{Proof:} We study both terms of
$\mathcal{C}_{3,\mathrm{sc}}$ separately.\\
\textit{Term 1}. Since $x$ is small we can use the approximation (\ref{A.19}) to
obtain
\begin{eqnarray}\label{A.21}
&&\hspace{-20pt} x \frac{2}{\sqrt{x}}\int^x_0 dy f(x-y) f(x) f(y)
\big(f(x)^{-1} - f(y)^{-1} - f(x-y)^{-1}\big)\nonumber\\
 &&\hspace{10pt}\cong (a_0)^3 b_0 x\frac{2}{\sqrt{x}} \int^x_0 dy
 ((x-y)yx)^{-1}(x^{3/2}-y^{3/2}-(x-y)^{3/2})\nonumber\\
 &&\hspace{10pt}= 2(a_0)^3 b_0 \int^1_0 dy
( (1-y)y)^{-1}(1-y^{3/2}-(1-y)^{3/2}) \,.
\end{eqnarray}
\textit{Term 2}. As before we split the $y$ integration into $[x,\delta]$ and
$[\delta,\infty)$. For the first integral we follow (\ref{A.21}) and
arrive at
\begin{eqnarray}\label{A.22}
&&\hspace{-20pt} x \frac{4}{\sqrt{x}}\int^\delta_x dy f(x)
f(y)f(y-x)
\big(f(y-x)^{-1} + f(x)^{-1} - f(y)^{-1}\big)\nonumber\\
 &&\hspace{10pt}\cong 4(a_0)^3 b_0 \int^{\delta/x}_1 dy
 (y(y-1))^{-1}\big((y-1)^{3/2}+1-y^{3/2}\big)\,,
\end{eqnarray}
which tends to the expression in (\ref{A.20}) as $x\to 0$.

For the second integral one has
\begin{eqnarray}\label{A.23}
&&\hspace{-20pt} 4 \sqrt{x} \int^\infty_\delta dy f(y-x) f(y) +4
f(x)\sqrt{x}\big(\int^\infty_\delta dy f(y) -
\int^\infty_\delta dy f(y-x)\big)\nonumber\\
 &&\hspace{10pt}= 4\sqrt{x} \int^\infty_\delta dy f(y)^2 + 4 f(x)
 \sqrt{x} x f(\delta)\,,
\end{eqnarray}
which is proportional to $\sqrt{x}$ and thus subleading.

\section{Appendix: Singular solutions for the Boltz\-mann-Nordheim equation}\label{sec.C}
 \setcounter{equation}{0}

We state in more detail the recent results of Escobedo, Mischler,
and Vel\'{a}squez \cite{EMV}. They consider the Boltzmann-Nordheim
equation (\ref{2.2}) with initial $f$ which diverges as
$x^{-7/6}$ for $x\to 0$. More precisely they assume that the initial
$f$ is once continuously differentiable on $(0,\infty)$ and
satisfies
\begin{equation}\label{C.1}
    |f(x)- A x^{-7/6}|\leq B x ^{-(7/6-\delta)}\,,\quad 0\leq x\leq 1\,,
\end{equation}
\begin{equation}\label{C.2}
    |f'(x)+\tfrac{7}{6} A x^{-13/6}| \leq B x ^{-(13/6-\delta)}\,,\quad 0\leq x\leq 1\,,
\end{equation}
\begin{equation}\label{C.3}
f(x)\leq B \mathrm{e}^{-D x}\,,\quad 1\leq x
\end{equation}
for some positive constants $A,B,D$, and $\delta$. Note that
$\rho(f)< \infty$, $\mathsf{e}(f)<\infty$ and that there is no
condition on their size. In particular $\rho(f)$ could be much
smaller than $\rho_\mathrm{c}$.\medskip\\
\textbf{Theorem} (EMV). \textit{For any $f$ satisfying (\ref{C.1}) to
(\ref{C.3}), there exists a unique solution, $f(x,t)$, to
(\ref{2.2}), continuously differentiable in $t$ for $0< t<\infty$
and continuous in $x$ for $0<x <\infty$, as well as a function
$b(t)$, satisfying
\begin{equation}\label{C.4}
0\leq f(x,t)\leq L x ^{-7/6}\mathrm{e}^{-D x}\,,\quad x> 0\,,\quad
t\in (0,T)\,,
\end{equation}
\begin{equation}\label{C.5}
    |f(x,t)- b(t)x^{-7/6} | \leq L x ^{-(7/6-\delta/2)}\,,\quad 0\leq x\leq
    1\,,\quad t\in(0,T)\,,
\end{equation}
\begin{equation}\label{C.6}
0\leq b(t)\leq L \,,\quad t\in (0,T)\,,
\end{equation}
for some positive constant $L$ and for some $T> 0$, $T$ depending on
$A,B$, and $\delta$.}\medskip

Together with our Proposition 1 it then follows that, for $0\leq t<
T$,
\begin{equation}\label{C.7}
\mathrm{e}(f(t))=\mathrm{e}(f)\,,
\end{equation}
\begin{equation}\label{C.8}
\rho(f(t))=\rho (f)-\Gamma_4 \int^t_0 ds b(s)^3\,,
\end{equation}
\begin{equation}\label{C.9}
n(t)=\Gamma_4 \int^t_0 ds b(s)^3\,.
\end{equation}
Thus the condensate density increases linearly in $t$ for small $t$.

It would be of interest to extend such a result to the coupled
system. As argued, we expect that $\mathcal{C}_3$ dominates
$\mathcal{C}_4$ and therefore in (\ref{C.5}) the divergence at $x=0$
should drop from $x^{-7/6}$ to $x^{-1}$, as also reported in
\cite{ST} for numerical solutions.

Currently, the restriction to bounded $T$ is needed for the proof.
Whether there is really such a restriction can only be speculated.
One could argue that by the choice of initial conditions one has
opened already a channel for the formation of condensate. If the
system uses only that channel, then
\begin{equation}\label{C.10}
    \Gamma_4 \int^\infty_0 dt b(t)^3 < \rho(f)
\end{equation}
and necessarily $b(t)\to 0$ as $t\to \infty$.  Once $f(t)$
is subcritical, the mass transfer could stop and $f(t)$ should converge
for large $t$ to the corresponding
unique equilibrium distribution.

\section{Appendix: Long time behavior}\label{sec.B}
 \setcounter{equation}{0}
 Lu \cite{Lu05} considers Equation (\ref{2.2}) in the weak form,
\begin{equation}\label{B.25c}
\frac{d}{dt}\int^\infty_0 \varphi(x) f(x,t)\sqrt{x}dx =
\int^\infty_0 \varphi(x)\mathcal{C}_4(f(t))(x) \sqrt{x}dx
\end{equation}
with test functions $\varphi\in C_\mathrm{b}^2(\mathbb{R}_+)$ and
proves that (\ref{B.25c}) remains meaningful even if
$\sqrt{x}f(x,t)dx$ is substituted a positive measure on
$\mathbb{R}_+$. We denote such a measure by $f(dx,t)$. The possible
roughness of $f(dx,t)$ is balanced by rewriting the right hand side
of (\ref{B.25c}) in such a way that the collisional difference
$\varphi(x) + \varphi(y+z-x) - \varphi(y) -\varphi(z)$ appears. For
$t=0$ we assume finite mass and energy, i.e.
$\int^\infty_0f(dx)<\infty$ and $\int^\infty_0x f(dx)<\infty$. Then
(\ref{B.25c}) has a solution $f(dx,t)$, which conserves mass and
energy. It is not known whether the solution constructed by Lu is
identical to the solution of the coupled system
(\ref{3.10}), (\ref{3.11}).
The stationary
measures of (\ref{B.25c}) are necessarily of the form
\begin{equation}\label{B.25d}
f_{\mathrm{tot},\beta,\mu}(dx)= f_{\beta,\mu}(x)
\sqrt{x} dx+n_\mathrm{con}\delta(x)dx
\end{equation}
with either $\mu\leq 0$ and condensate density $n_\mathrm{con} = 0$ or $\mu=0$ and
$n_\mathrm{con} > 0$.

Let us assume that the initial data are given by a density, i.e.
$f(dx)=f_{\mathrm{ac}}(x)\sqrt{x} dx$, such that
\begin{equation}\label{B.25e}
\rho=\rho(f)=\int^\infty_0
f_{\mathrm{ac}}(x)\sqrt{x}
dx<\infty\,,\quad
\mathsf{e}=\mathsf{e}(f)=\int^\infty_0
x f_{\mathrm{ac}}(x)\sqrt{x}
dx<\infty\,.
\end{equation}
In general, the solution to (\ref{B.25c}) is then a measure,
$f(dx,t)$. It can be uniquely decomposed into an absolutely
continuous and a singular part,
\begin{equation}\label{B.30}
f(dx,t)=f_\mathrm{ac}(x,t)\sqrt{x}dx
+ f_\mathrm{s}(dx,t)\,.
\end{equation}
Now let $f_{\beta,\mu}$ be the equilibrium distribution
corresponding to $\rho(f)$, $\mathsf{e}(f)$, where $\mu=0$ in case
$n_{\mathrm{con}} = \rho - \rho_\mathrm{c} > 0$.  Lu \cite{Lu05} proves that, for
$\rho\leq\rho_\mathrm{c}$,
\begin{equation}\label{B.31}
\lim_{t\to\infty} \int^\infty_0
f_\mathrm{s}(dx,t)=0\,,\quad \lim_{t\to\infty}
\int^\infty_0
|f_\mathrm{ac}(x,t)-f_{\beta,\mu}(x)|
\sqrt{x}dx=0\,.
\end{equation}
On the other hand for $\rho >\rho_\mathrm{c}$ one has
\begin{equation}\label{B.31a}
\lim_{t\to\infty} \int^\infty_0
|f_\mathrm{ac}(x,t)-f_{\beta,0}(x)|
x\sqrt{x}dx =0
\end{equation}
and
\begin{equation}\label{B.32}
\lim_{t\to\infty} \int_{\{x|\,0 \leq x^4\leq r(t)\}}
f(dx,t)=\rho - \rho_\mathrm{c}\,,
\end{equation}
where $r(t)$ is the integral in (\ref{B.31a}). Thus for large times
the solution has a $\delta$-peak like concentration in the interval
$[0,r(t)^{1/4}]$ with weight $n_\mathrm{con}$.\bigskip\\

\end{appendix}

\noindent \textbf{Acknowledgements}. I am grateful to Oliver Penrose for
stimulating my interest in the kinetics of the Bose-Einstein
transition. My thanks go to Xuguang Lu for extensive email exchanges
which helped me in understanding the transition mechanics.

\end{document}